\documentclass[twocolumns]{aa} 

\usepackage{graphicx}
\usepackage{txfonts}
\usepackage{hyperref}
\begin{document} 
    \title{The dynamical age of the LMC globular cluster NGC~1835 using
      the ``dynamical clock''\thanks{Based on observations with the
    NASA/ESA HST, obtained under program GO 16361 (PI: Ferraro). The
    Space Telescope Science Institute is operated by AURA, Inc., under
    NASA contract NAS5-26555}} \subtitle{Star density profile and blue
      straggler stars}

   \author{Camilla Giusti
          \inst{1}
          \inst{2}
          \and
          Mario Cadelano
          \inst{1}
          \inst{2}
          \and
          Francesco R. Ferraro
          \inst{1}
          \inst{2}
          \and
          Barbara Lanzoni
          \inst{1}
          \inst{2}
          \and
          Cristina Pallanca
          \inst{1}
          \inst{2}
          \and
          Enrico Vesperini
          \inst{3}
          \and
          Emanuele Dalessandro
          \inst{2}
          \and
          Maurizio Salaris
          \inst{3}
          }

\institute{Dipartimento di Fisica \& Astronomia, Universit\`a degli Studi di Bologna, via Gobetti 93/2, I-40129 Bologna, Italy\\
\and
INAF - Astrophysics and Space Science Observatory Bologna, Via Gobetti 93/3, 40129, Bologna, Italy \\
\and 
Dept. of Astronomy, Indiana University, Bloomington, IN 47401, USA\\
\and 
Astrophysics Research Institute, Liverpool John Moores University, Liverpool L3 5RF, UK \\
}


 
\abstract {In the context of the study of the size-age relationship observed in star clusters in the Large Magellanic Cloud and the investigation of its origin, here we present the determination of the structural parameters and
  the dynamical age of the massive cluster NGC 1835.  We have used a
  powerful combination of optical and near-ultraviolet images acquired
  with the WFC3 onboard the HST to construct the star density profile
  from resolved star counts, determining the values of the core,
  half-mass and tidal radii through the comparison with the King model
  family.  The same data also allowed us to evaluate the dynamical age
  of the cluster by using the `dynamical clock'.  This is an
  empirical method that quantifies the level of central segregation of
  blue stragglers stars (BSSs) within the cluster half-mass radius by
  means of the $A^+_{rh}$ parameter, which is defined as the area
  enclosed between the cumulative radial distribution of BSSs and that
  of a reference (lighter) population.  The results confirm that NGC
  1835 is a very compact cluster with a core radius of only 0.84
  pc. The estimated value of $A^+_{rh}$ ($0.30\pm 0.04$) is the
  largest measured so far in the LMC clusters, providing evidence of a
  highly dynamically evolved stellar system.  NGC 1835 nicely fits
  into the correlation between $A^+_{rh}$ and the central relaxation
  time and in the anti-correlation between $A^+_{rh}$ and the core
  radius defined by the Galactic and the Magellanic Cloud clusters  investigated to date.}


   \maketitle
%
\section{Introduction}
\label{intro}
The Large Magellanic Cloud (LMC) hosts globular clusters
  (GCs) spanning various ages, ranging from a few million to billions
of years \citep{bica+2008, nayak+2016}. Its star formation history
differs significantly from that of the Milky Way (MW), which
predominantly hosts very old ($t>10$ Gyr) GCs. As suggested,
  e.g., by Fig. 15 of \citet{mackey+2003} and Fig. 1 of
  \citet{ferraro+2019}, during the initial stages of star formation in
  the LMC (around 13 Gyrs ago), massive clusters (M > $10^5
  M_{\odot}$) formed at various distances from the galaxy centre. This
  phase was followed by a period of quiescence, known as the
  `age-gap', lasting about 10 billion years. Then, approximately 3 Gyr
  ago, a period of strong tidal interaction between the MW and the
  LMC, and/or between the LMC and the Small Magellanic Cloud (SMC)
  likely began, triggering significant collisions among gas clouds
  that boosted a second burst of star formation, during which the less
  massive clusters now observed in the LMC's central regions have been
  generated \citep{dacosta+1991, rich+2001, bekki+2004, mazzi+2021}.

Unlike the MW, the LMC provides an opportunity to explore the
properties of GCs spanning a large range of ages and masses \citep{elson+1989, elson+1991, 
olszewski+1996, olsen+1998, brocato+1996,
  mackey+2003,baumgardt+2013,ferraro+1995,
  ferraro+2004, mucciarelli+2006}. From
these studies, about 30 years ago emerged a peculiar trend between the
core radius ($r_c$) and the chronological (stellar) age of the
clusters, the so-called `size-age conundrum'. Specifically,
while young clusters all have small core radii ($r_c$<2.5 pc), the
older ones exhibit a broader range of sizes, reaching up to $r_c=10$
pc \citep{elson+1989, elson+1991, mackey+2003}. Different solutions to
the size-age conundrum have been proposed in the literature, with the
most acknowledged one being that presented in
\citet{mackey+2008}. These authors suggest an evolutionary connection
between the younger and older clusters, with the former representing
how the latter appeared at the epoch of their formation. Hence, in
this framework all clusters were born compact, and subsequent
interactions among single stars and binary stellar-mass black holes
(BHs) drove core expansion up to the large values of $r_c$ currently
observed for the oldest systems.\\  However, \citet{ferraro+2019} put
into evidence a significant difference between the young and the old
LMC clusters, and challenged such an evolutionary connection; specifically, \citet{ferraro+2019} pointed out that all the young clusters are
significantly less massive than the old ones (see the middle panel in
Fig. 1 of \citealt{ferraro+2019}). \citet{ferraro+2019}
proposed instead a new hypothesis to explain the size-age conundrum where the
differences in the core radius values among the oldest systems are
attributed to variations in their dynamical ages. In fact, GCs are
collisional systems, in which the interactions among stars
can alter significantly the internal energy budget. Because of these
interactions, the heavier stars tend to progressively migrate towards
the central regions through the action of dynamical friction, causing a contraction of the
core and an increase in the central density until this collapse is halted by the energy provided by interactions with primordial or dynamically formed binary stars \citep{heggie+2003}. The evolutionary path of a cluster depends both on its internal structure and the tidal field of its host galaxy (see, e.g., \citealp{meylan_heggie97}). For instance, clusters with higher central density experience
  a more rapid dynamical evolution than others, due to a larger
  probability of stellar interactions. Similarly, in clusters located
  at small galactocentric distances, the stripping action of low-mass
  stars by the galactic tidal field accelerates the process.  This
implies that clusters of the same chronological age may be at
different stages of their dynamical evolution depending on differences
in their initial structural properties and the external environment
where they evolved. In particular, in the scenario proposed to explain
the observed variety of core radii for the old GCs, loose systems
represent dynamically young systems, while the compact ones are the
systems in more advanced stages of their dynamical evolution.

\citet{ferraro+2019} demonstrated this hypothesis by measuring the
dynamical ages of five old and coeval clusters in the LMC through the
`dynamical clock' method. This approach relies on the
properties of a distinct stellar population routinely observed within
GCs and known as blue straggler stars (BSSs). The formation processes of these
  peculiar objects are not yet fully understood, but two main
  formation scenarios have been proposed so far (see also
  \citealt{bailyn95}): i) mass-transfer in binary systems
  \citep{mccrea+1964}, and ii) stellar mergers due to direct
  collisions between two or more stars \citep{hills+1976}. These are
  both mass-increasing processes, making BSSs more massive (M$\sim 1
- 1.4 M_\odot$, \citealt{fiorentino+2014, raso+2019}) than
the average of normal stars in old clusters (M$\sim 0.3-0.4
M_\odot$). Therefore, under the action of dynamical friction, they
sink towards the central region faster than less massive stars. For
this reason, they act as excellent gravitational probes and their
degree of central segregation compared to that of a normal cluster
population can be used as a clock-hand of a `dynamical chronometer' to
measure the dynamical stage of a system (see \citet{ferraro+2012,
  ferraro+2018, ferraro+2019, ferraro+2020, ferraro+2023,
  lanzoni+2016}. In order to quantify the level of BSS central
segregation, \citet{alessandrini+2016} (see also
\citealt{lanzoni+2016}) introduced the $A^+_{rh}$ parameter. It is
defined as the area enclosed between the cumulative radial
distribution of BSSs and that of a reference population (lighter
stars, like main sequence or red giant branch stars) within one
half-mass radius ($r_h$) from the cluster centre. The value of
$A^+_{rh}$ provides a measure of the degree of mass segregation of BSS
and generally increases during a cluster dynamical evolution (see also
\citet{alessandrini+2016} for a discussion of the effects of dark
remnants on the evolution of BSS segregation and the $A^+_{rh}$
parameter). This has been demonstrated by the strong correlation
between $A^+_{rh}$ and the number of central relaxation times
undergone by the system over its lifetime ($N_{\rm relax}$) in a
sample of about 50 Galactic GCs \citep{ferraro+2018,
  ferraro+2023,beccari+2023}.  By applying the dynamical clock to five
old clusters in the LMC (namely, NGC 1466, NGC 1841, NGC 2210, NGC
2257, Hodge 11), \citet{ferraro+2019} found that they follow the same
correlation between $N_{relax}$ and $A^+_{rh}$ drawn by the Milky Way
systems, thus demonstrating the validity of the method beyond
our Galaxy. Furthermore, this study proved the expected
anticorrelation between the core radius and the dynamical age of the
five systems, confirming that the wide range of core radii observed
for the old LMC clusters is due to their different dynamical stage.

The cluster sample surveyed so far, however, leaves the oldest LMC
systems with a very compact core radius still totally unexplored. To
fill this gap, here we apply the dynamical clock method to NGC
1835. This cluster is very old (approximately $12.5$ Gyr old;
\citealt{giusti+2024,olsen+1998}) and very compact ($r_c< 0.8$ pc),
with a large mass (approximately $6 \times 10^5 M_\odot$;
\citealt{mackey+2003}) and a low metallicity ([Fe/H]$\sim -1.7$ dex;
\citealt{mucciarelli+2021}).  The present analysis is based on a set
of high-resolution Hubble Space Telescope (HST) images acquired with
the WFC3 in the cluster direction. These exposures were already used
to characterize the properties of a small stellar system named KMK
88-10 located at only $2\arcmin$ from the cluster, providing evidence
that it has been possibly captured by NGC 1835 and it is on the verge
of tidal distruption \citep{giusti+2023}. Moreover, in
\citealt{giusti+2024} we presented the discovery of a very extended
blue tail in the horizontal branch (HB) of NGC 1835, and we provided
the most precise estimate so far of the cluster age ($12.5\pm 1$ Gyr),
confirming that it is very old.  In this paper we determine the star
density profile of the cluster to characterise its structural
parameters and to investigate its possible core-collapse nature, as
indicated in \citet{mackey+2003}, and we measure its
dynamical age by using the level of BSS segregation, discussing the
results in the context of the size-age conundrum.

\section{Data analysis}
\label{dataanalysis}
The present study is based on high-resolution images collected with
the UVIS channel of the HST/WFC3 in the F300X, F606W, F814W filters,
complemented with a simultaneous parallel observations in the F606W
and F814W filters acquired with the Wide Field Camera of the Advanced
Camera for Surveys (ACS/WFC), under program GO 16361 (PI: Ferraro).
The data set and data reduction procedures are described in
\citet{giusti+2024}. Here we briefly summarize the main points.

The WFC3 has a total field of view of $160\arcsec \times 160\arcsec$,
and the cluster centre has been located at the UVIS1 aperture. These data therefore
sample the cluster population, both in the optical (F606W and F814W)
and in the near-ultraviolet (near UV) band (F300X). The parallel ACS
observations sample an LMC field located at $\sim5\arcmin$ from the
WFC3 pointing, and are therefore used for decontamination purposes.
The data reduction procedure was executed with the DAOPHOT II
software, following the methodologies delineated in earlier
publications (see, e.g. \citealt{cadelano+2022a, onorato+2023, chen+2023}).
The only peculiarity is that for the WFC3 data set we followed the
so-called `UV-guided search', which is specifically designed to
maximize the detection of hot stars (like extremely-blue HB stars,
BSSs, and white dwarfs) in stellar populations where giant stars are
dominant (see \citealt{paresce+1992, ferraro+1997, ferraro+1998, ferraro+1999,
  ferraro+2001, ferraro+2003a, lanzoni+2007, dalessandro+2013,
  raso+2017,chen+2021,chen+2022}).  This approach has led to the
discovery of the extended HB blue tail in NGC 1835
\citep{giusti+2024}, which remained undetected in all the previous
optical studies of the cluster, and plays a crucial role in the
present paper, where the collection of a complete sample of BSSs is
required.  The approach consists in generating a master list of stars
identified in at least half of the near UV (F300X) images, and then
force fitting the PSF model to the position of these stars in all the
other filters.  The magnitude values estimated for each star from
different images have been combined using DAOMATCH and DAOMASTER, and
finally calibrated onto the VEGAMAG photometric system by applying the
aperture corrections and zero points quoted in the dedicated HST web
pages.  The instrumental positions have been corrected for geometric
distortions (see \citealp{bellini+2011} and \citealt{meurer+2003} for
the WFC3 and the ACS, respectively) and then transformed into absolute
coordinates through cross-correlation with the Gaia DR3 catalog
\citep{gaiacollaboration+2022}.

\begin{figure*}
     \centering
         \includegraphics[scale = 0.48]{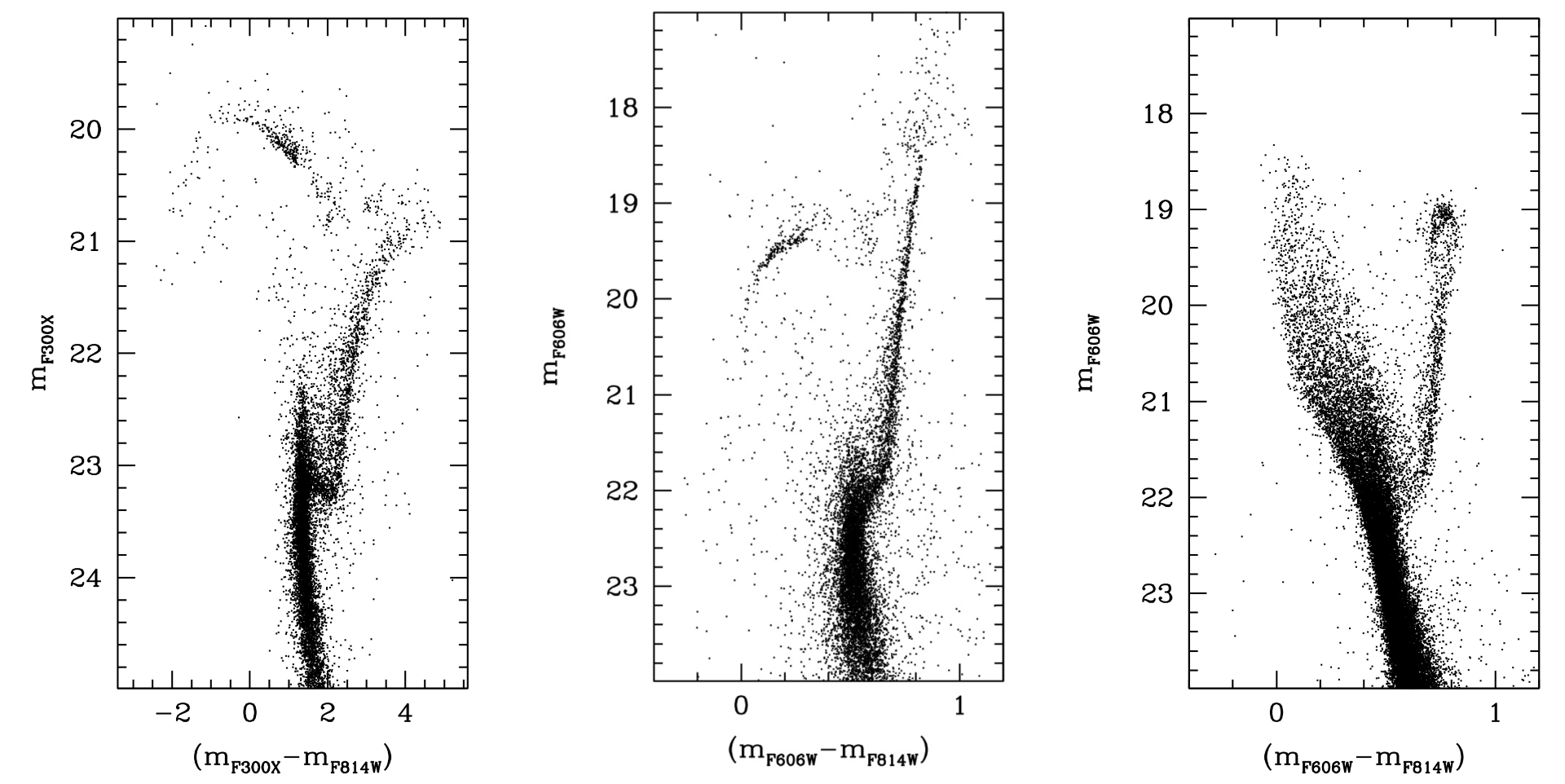}
         \caption{CMDs of the cluster region and the field region in different filter combinations. \emph{Left panel}: near UV
         CMD of NGC 1835 for the  stars sampled within
         $r<13\arcsec$ from the centre. \emph{Central panel}: Optical
         CMD of NGC
         1835 for the same stars as in the left panel.
         \emph{Right panel}: Optical
         CMD of the LMC field region sampled by the ACS parallel
         observations.}
         \label{cmds}
\end{figure*}

The left-hand and central panels of Figure \ref{cmds} show the near UV
and the optical CMDs of NGC 1835, respectively: only stars within the
innermost $13\arcsec$ (corresponding to the cluster half mass radius,
see Section \ref{kingdens}) are plotted to more clearly highlight the
cluster population. Being stars with high temperatures ($T_{\rm
  eff}\sim 7000$ K), BSSs are better distinguishable in the near UV
plane, where they span a range of about 2.5 magnitudes ($20.5<m_{\rm
  F300X}<23$) above the main sequence turn-off (MS-TO) region.
On the other hand, the cold red giant branch (RGB) sequence is better
delineated in the optical CMD (at $17.5<m_{\rm F814W}<21.5$). For
these reasons, the BSS sample and the RGB reference population used in
the following analysis (see Section \ref{BSS}) will be selected in the
near UV and optical CMDs, respectively. In the right-hand panel we
show the CMD of the LMC field region sampled by the ACS parellel
observations, in the same filter combination of the cluster optical
CMD.

\section{Gravitational centre}
\label{sec:centre}
As already discussed in many works \citep[e.g.][]{ferraro+1999,
  ferraro+2003b, lanzoni+2007, ibata+2009, miocchi+2013}, the precise location of
the centre of gravity plays a very important role in the accurate
determination of the star density and surface brightness
profiles. \citet{mackey+2003} determined the centre position of NGC
1835 from the location of the cluster's surface brightness peak,
finding: $\alpha = 05^h:05^m:06^s.7$, $\delta=
-69^{\circ}:24\arcmin:15\arcsec$. However, this method is prone to
significant biases due to the possible presence of even a few bright
stars not located at the cluster's centre, because the surface
brightness peak would be systematically shifted towards them. Thanks
to our high-resolution dataset, we have been able, instead, to
determine the position of the gravitational centre ($C_{\rm grav}$)
from resolved star counts using the method outlined in
\citet{montegriffo+1995} (see also \citet{lanzoni+2007,
  lanzoni+2019}).

This consists of an iterative procedure that starts from a test value
of the centre (e.g. a literature value) and determines the average of
the coordinates of a sample of stars within a fixed distance $r$. This
procedure produces a new estimate of the centre, from which the next
iteration starts. The convergence is fixed when the difference between
the position of two successive centres is less than 0.01\arcsec.  The
search radius $r$ must be chosen carefully. It should be larger than
the core radius $r_c$ ($4.76\arcsec$ for NGC 1835;
\citealt{mackey+2003}), to work in a region where the stellar density
is not uniform.
On the other hand, excessively large radii lead to decreased
sensitivity to the central concentration. The procedure also allows
for a magnitude selection of the sample stars. A faint magnitude limit
is necessary to obtain sufficient statistics, but a too high value
introduces spurious incompleteness effects. In consideration of these
compromises, we iterated the process considering different radius
values ($r \le 10\arcsec, 15\arcsec, 20$\arcsec) and different
magnitude limits ($m_{\rm F814W} < 21.5, 22, 22.5$), ensuring in each
combination a minimum of 70\% completeness (see Sec.\ref{apiu}) and thousands of stars
included within the search radius. We repeated the iterative procedure
for the nine possible combinations. The final value of $C_{\rm grav}$
was then obtained from the average of these nine values:
$\alpha=05^h:05^m:6^s.71$,
$\delta=-69^{\circ}:24\arcmin:14\arcsec.78$.  Their dispersion
provided us with an uncertainty $\sigma = 0.25\arcsec$. Our new
estimate of $C_{\rm grav}$ is consistent within 1$\sigma$ with that
quoted by \citet{mackey+2003}.

\section{Star count density profile}
\label{sec:densprof}
We constructed the projected density profile of NGC 1835 from resolved
star counts following the procedure described in \citet{miocchi+2013,
  lanzoni+2019, raso+2020}. We used only evolved stars ($m_{\rm
  F300X}<22.5$), in order to work with objects of approximately the
same mass and to ensure a completeness of $\sim 80$\% in the central
region.  We also excluded the brightest sources ($m_{\rm F300X}<18$)
to avoid possible biases due to saturation. We divided the WFC3 field
of view into 18 radial rings centered on the cluster centre (see
Sec. \ref{sec:centre}) and we divided each ring into four
sub-sectors. For each sub-sector, we counted the number of stars in it
and we divided the result by the sector's area. We adopted as the
density value in each ring the average of the values in the four
corresponding sub-sectors, while the error was calculated as their
standard deviation. The resulting projected density profile,
$\Sigma_*(r)$, is shown in the left panel of Figure \ref{profili}
(empty circles). The profile shows a central plateau out to $\sim 3
\arcsec$, followed by a gradually decrease in stellar density as the
distance from the centre increases.
A centrally-flat profile suggests that the cluster is not
  core-collapsed yet, since a steep power-law cusp in the innermost
  portion of the density distribution is expected for systems that
  already experienced this process \citep[see,
    e.g.,][]{meylan_heggie97}.  In the outer regions ($r>40\arcsec$),
where the LMC field contribution becomes predominant over that of the
cluster, a constant density value is observed. From averaging the
values of the last 4 bins we derived the LMC mean field density
($\log\Sigma_{*}\sim-0.9$; see the horizontal dashed line in
Fig. \ref{profili}). We subtracted this quantity from the density
value measured in each bin and we then obtained the background
decontaminated profile, which is shown as filled circles in the left
panel of the figure.

\subsection{Fit of the Density Profile}
\label{kingdens}
To determine the structural parameters of NGC 1835, we determined the
best-fit model to the cluster's background decontaminated density
profile by comparing it with the family of spherical, isotropic,
single-mass King models \citep{king+1966}. The single-mass assumption
works with our observed density profile since it was constructed by
selecting a sample of stars of approximately the same mass. We
followed the procedure outlined in \citet{raso+2020} (see also
\citet{pallanca+2023, deras+2023, deras+2024}). It consists in a Monte Carlo
Markov Chain (MCMC) fitting technique assuming flat priors for the
fitting parameters (namely the central density, the concentration
parameter $c$, and the core radius $r_c$) and a $\chi^2$
likelihood. The resulting best-fit model is represented by the red
line in the left panel of Fig. \ref{profili}.\\ From this procedure we
find that NGC 1835 is a cluster characterized by a relatively high
concentration parameter ($c=1.47^{+0.15}_{-0.14}$, corresponding to a
  dimensionless central potential $W_0=6.8\pm0.5$), a very compact core
radius $r_c = 3.5^{+0.7}_{-0.5} \, \text{arcsec}$, a half-mass radius $r_h = 13^{+3}_{-1} \, \text{arcsec}$, and a tidal radius $r_t = 109^{+36}_{-18} \, \text{arcsec}$. Knowing the LMC distance
 (49.6 kpc,
\citealt{pietrzynski+2019}) the results can be expressed as $r_c= 0.84$
pc, $r_h=3.13$ pc, and $r_t= 26.23$ pc (see Table
\ref{table:1}). \citet{mackey+2003} had previously determined some of
the structural parameters of the cluster by fitting the observed
surface brightness profile with an EFF model \citep{elson+1987}. In
particular, they find $r_c = 4.76\arcsec$, the small difference
with respect to our result being likely attributable to the different
adopted approaches.  The residuals of the fit (see bottom panel of
Fig. \ref{profili}) show no significant discrepancy from the trend
expected from a King model profile in the central regions, further
confirming that NGC 1835 is not a core collapse cluster. This is in
disagreement with the work of \citet{mackey+2003}, where NGC 1835 was
classified as a possible core collapse system since a low significance
cusp was detected.

\subsection{Surface Brightness Profile}
\label{sbprof}
\begin{figure*}
     \centering
         \includegraphics[scale = 0.32]{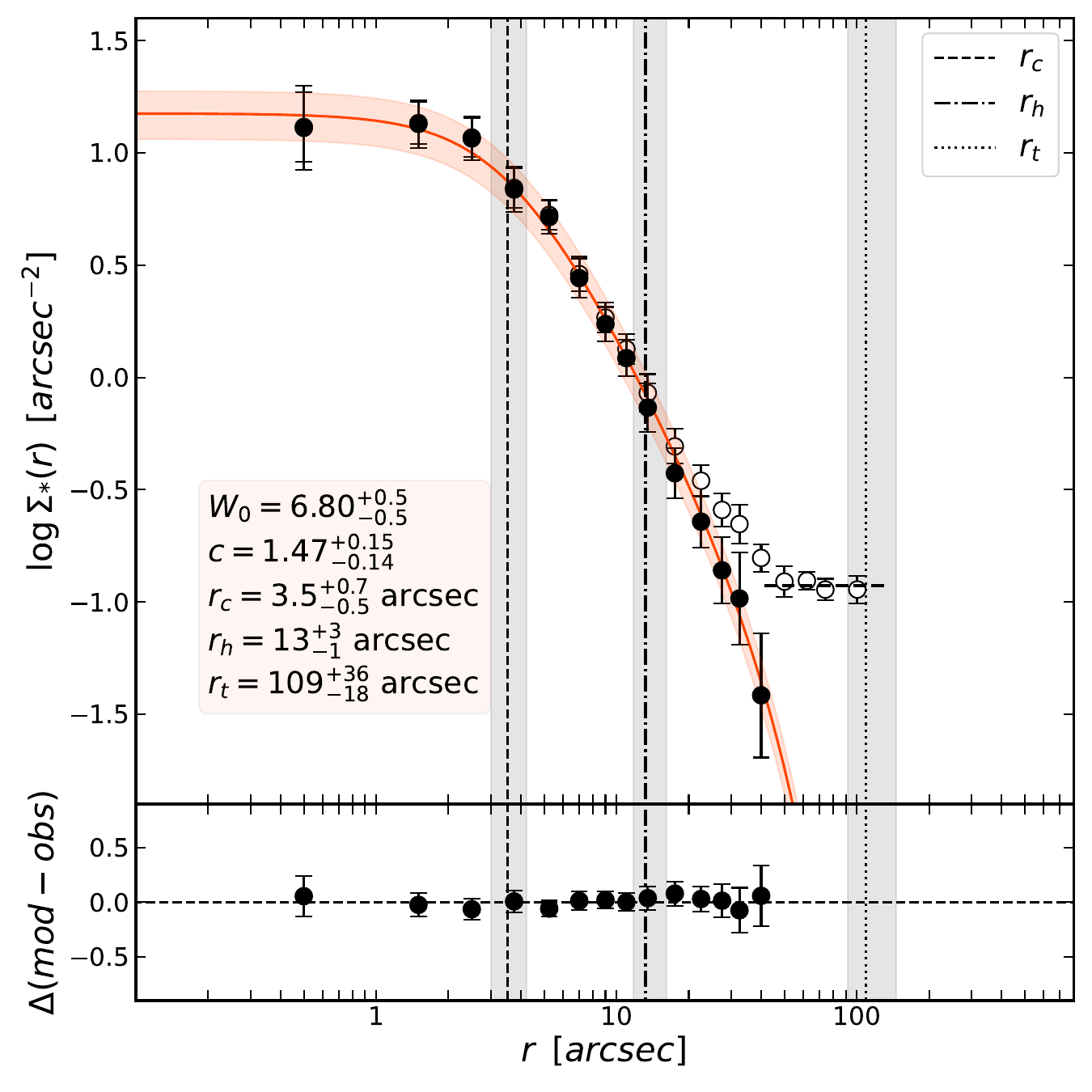}
         \quad
         \includegraphics[scale = 0.32]{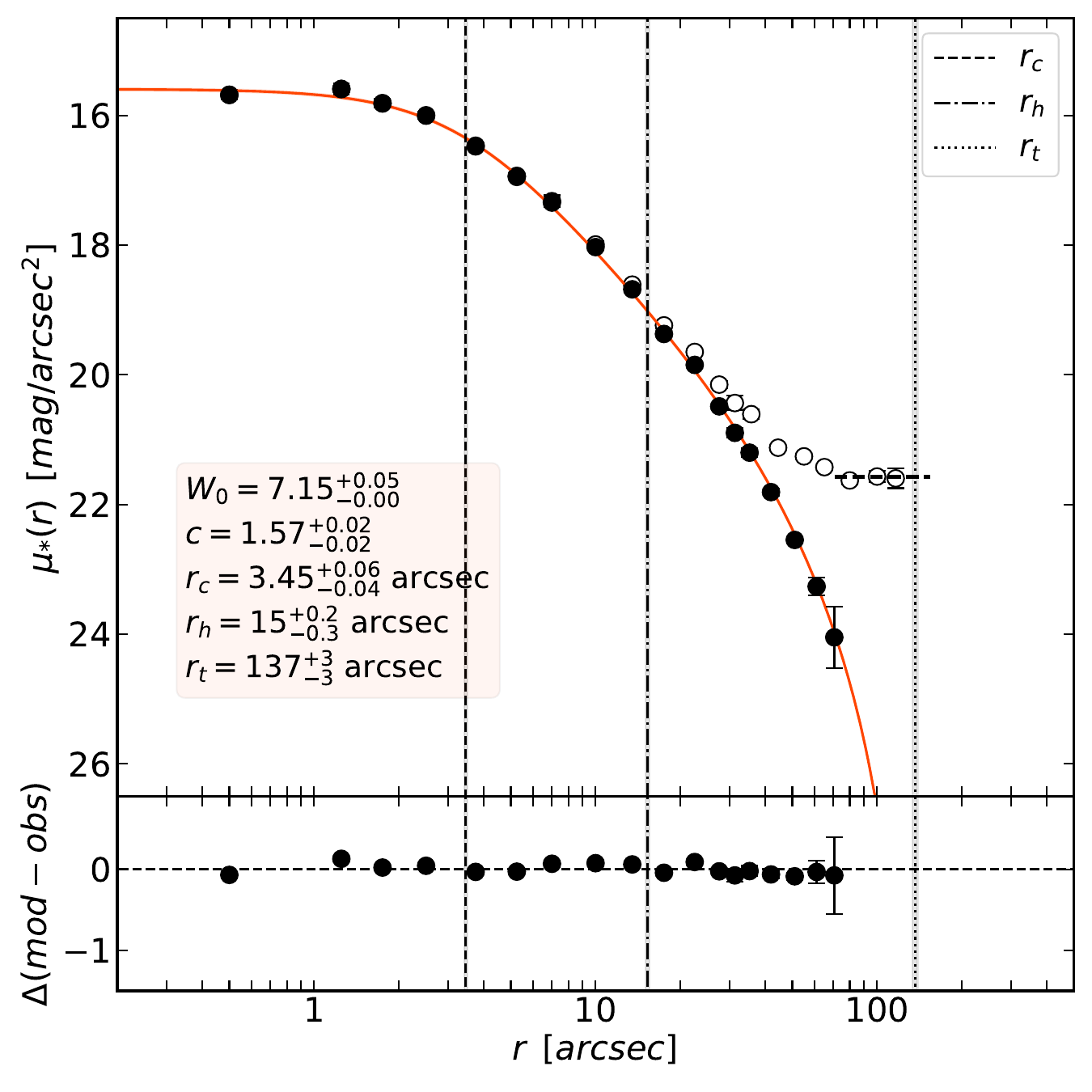}
         \caption{Density and brightness profiles of NGC 1835. \emph{Left panel}: Projected density profile of NGC
           1835 obtained from star counts in concentric annuli around
           the cluster centre (empty circles). The horizontal dashed
           line marks the LMC field density that has been subtracted
           to the observed points to determine the background
           subtracted profile (filled circles).  The red solid line
           represents the best-fit King model to the cluster density
           profile and the red stripe outlines the range of $\pm
           1\sigma$ solutions.  The vertical lines mark the locations
           of the core radius (dashed line), half-mass radius
           (dot-dashed line), and tidal radius (dotted line), with
           their respective 1$\sigma$ uncertainties indicated by gray
           stripes.  The values of the main structural parameters
           determined from the fitting process are also marked (see
           details in the text). \emph{Right panel}: Same as in the
           left panel, but for the surface brightness profile of NGC
           1835 in the F606W filter.}
\label{profili}
\end{figure*}

To obtain further confirmation of the cluster's structural parameters,
we also analyzed the surface brightness profile. The procedure is
analogous to the one adopted for the density profile (see
Sec. \ref{sec:densprof}). Also in this case, we divided the WFC3 field
of view into several concentric rings (21 rings from $C_{\rm grav}$ up
to $\sim 120 \arcsec$) and each ring into four sub-sectors. We adopted
as integrated surface brightness of each ring the average of the four
values of the corresponding sub-sectors and we assumed the standard
deviation as its uncertainty. The resulting surface brightness profile
obtained in the $m_{\rm F606W}$ filter is shown in the right panel of
Fig. \ref{profili} (open circles). The average surface brightness of
the last 3 points provided us with the mean surface brightness of the
LMC background ($\mu_{\rm F606W} \sim 21.5$ mag arcsec$^{-2}$).
We subtracted this value from all the observed points, obtaining the
background decontaminated profile (solid circles). The fit to the
profile was carried out following the same procedure described above,
comparing it with the family of King models. The results have provided
a King concentration index $c=1.57\pm 0.02$, a central surface brightness
$\mu_{\rm F606W,0} = 15.6$ mag arcsec$^{-2}$, a core radius $r_c=3.45^{+0.06}_{-0.04}
\, \text{arcsec}$ (0.83 pc), an half-mass radius $r_h= 15^{+0.2}_{-0.3}\, \text{arcsec}$ (3.61 pc),
and a tidal radius $r_t= 137\pm3\, \text{arcsec}$ (32.72 pc). All these quantities
are in good agreement within the errors with the results obtained
from the fit of the star count density profile. Even in this case, no
evidence of central cusp is found.

\section{Measuring the dynamical age}
\label{BSS} 
This work is mainly devoted to determine the dynamical age of NGC 1835
through the dynamical clock, which measures the level of central
sedimentation of BSSs with respect to a lighter reference population,
using the $A^+_{rh}$ parameter \citep{alessandrini+2016}.  Hence,
after estimating the central relaxation time through the standard
approximated approach described in \citet{djorgovski+1993}, in this
section we discuss the necessary preliminary steps and the final
determination of $A^+_{rh}$.

\subsection{The central relaxation time}
\label{timerelax}
Under the assumption of a spherically symmetric and isotropic stellar
system well reproduced by a King model, we computed the central
relaxation time ($t_{rc}$) of the cluster following equation (10) in
\citet[][see also \citealt{spitzer+1987}]{djorgovski+1993}:
\begin{equation}
    t_{rc}=1.491\times10^7 \times \frac{k}{\ln(0.4N_{*})m_{*}}\rho_{M,0}^{1/2}r_c^3,
\end{equation}
where the constant $k\sim0.5592$, $N_*=M_{cl}/m_{*}$ is the number of
stars in the cluster of mass $M_{cl}$, $m_{*}=0.3 M_\odot$ is the
average stellar mass, $\rho_{M,0}$ is the central mass density in
units of $M_\odot$ pc$^{-3}$, $r_c$ is the core radius in parsecs, and $t_{rc}$ is expressed in years.  We adopted the cluster
total mass provided in \citep{mackey+2003} ($\log M/M_\odot = 5.83$
that was estimated by using a mass-to-light ratio $M/L_V = 3.56$)
rescaled to $M/L_{V}=2$, which is the typical value for 10-13 Gyr old
clusters \citep{maraston+1998}.  In addition, to keep the work
consistent with the assumptions of \citet{ferraro+2019}, we measured
$\rho_{M,0}$ from the equations (4), (5), (6) of
\citet{djorgovski+1993}, adopting the concentration parameter $c$ and
the core radius $r_c$ determined from the star density profile fit
(see Table \ref{table:1}), and the central surface brightness
$\mu_{555}(0) = 16.37$ quoted in \citet{mackey+2003}.
The result for NGC 1835 is $\log(t_{rc}/{\rm yr}) = 8.16$.

We used this value to obtain a first indication about the dynamical
status of the cluster through the parameter $N_{\rm relax}$. This is
defined as the ratio between the system's chronological age
and its current central relaxation time, thus quantifying the number
of $t_{rc}$ experienced by the system during its lifetime. A high
value of $N_{\rm relax}$ identifies dynamically evolved clusters,
whereas a low value corresponds to dynamically young systems. For NGC
1835, using the chronological age by \citet{giusti+2024} (12.5 Gyr) 
  we find $N_{\rm relax} = 86.9$.   This is one of the largest value
  obtained so far in the LMC (see \citet{ferraro+2023}), meaning that NGC 1835 is highly
  dynamically evolved.

\begin{table}[h!]\label{table}
\centering
\begin{tabular}{l l}
\hline
\hline
Parameter & Estimated value \\
\hline
Centre of gravity & $\alpha_{J2000}$ = $05^h05^m06^s.7$\\
 & $\delta_{J2000}$ = $-69^{\circ}24' 15''$ \\
Age &  $t=12.5 \pm 1$ Gyr \\
King concentration & $c=1.47^{+0.15}_{-0.14}$\\
Central dimensionless potential & $W_0=6.8^{+0.5}_{-0.5}$ \\
Core radius & $r_c = 3.5^{+0.7}_{-0.5}$ arcsec (0.84 pc)\\
Half mass radius & $r_h = 13^{+3}_{-1}$ arcsec (3.13 pc)\\
Tidal radius & $r_t = 109^{+36}_{-18}$ arcsec (26.23 pc)\\
Central relaxation time  & $\log t_{rc}=8.158$ (yr)\\
Age/Central relaxation time & $N_{\rm relax} = 86.9$\\
\hline
\end{tabular}
\caption{Summary of the main properties of NGC 1835 estimated in this
  work from the fit of the star counts density profile (see left panel of Fig.\ref{profili}). The age was estimated in \citet{giusti+2024}.}
\label{table:1}
\end{table}

\subsection{Selection of BSSs and reference population}
\label{bss_sel}

The first step for estimating the $A^+_{rh}$ parameter of NGC 1835 is
to select its population of BSSs and reference stars. In a cluster's
CMD, BSSs are found in the region bluer and brighter than the cluster
MS-TO. We followed the selection procedure already described in
several works (see \citealt{ferraro+1992, ferraro+2001, ferraro+2003a,
  ferraro+2023}.

\begin{figure*}
     \centering
         \includegraphics[scale = 0.38]{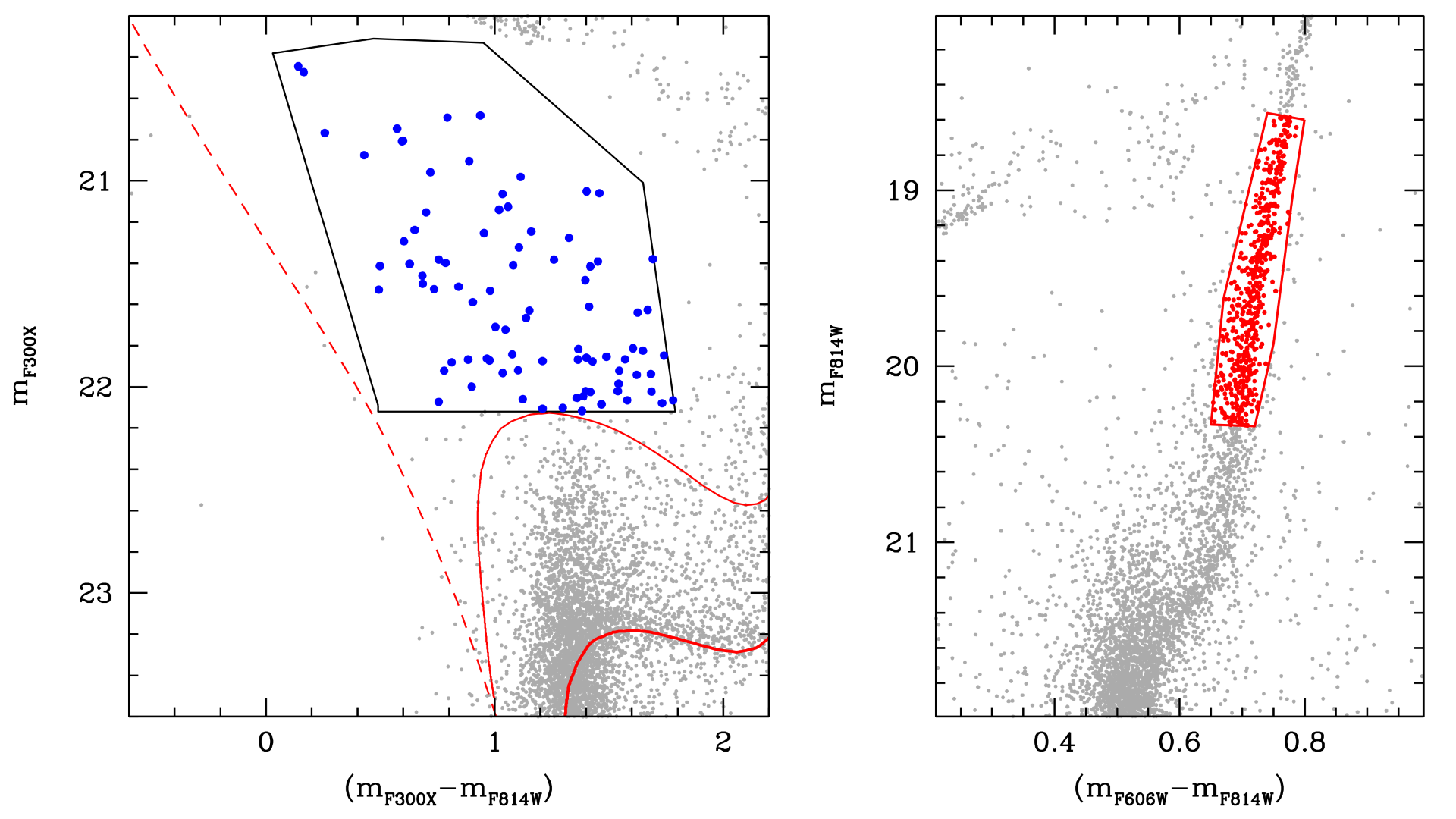}
         \caption{BSS and RGB samples selection. \emph{Left panel}: near UV CMD of NGC 1835 with the
           selected BSS sample shown with blue circles. The rightmost
           solid line is the evolutionary track corresponding to a
           mass $0.8 M_{\odot}$. The leftmost solid line is the
           evolutionary track of a $1 M_{\odot}$ mass. The dotted
           line is the PARSEC isochrone corresponding to a very young
           age (40 Myr), representing the cluster ZAMS. \emph{Right
           panel}: Optical CMD of NGC 1835 with the selected RGB
           sample shown with red circles. }
         \label{boxes}
\end{figure*}

Due to effective temperatures larger than those of MS and RGB stars,
BSSs are best selected in CMDs built with blue and possibly UV
filters.  We therefore drew the BSS selection box in the ($m_{\rm
  F300X},m_{\rm F300X}$-$m_{\rm F814W}$) UV plane: see the multi-sided
polygon shown in the left panel of Figure \ref{boxes}. We constructed
the left side of the polygon by adopting as an approximate reference
the cluster zero-age MS (ZAMS), represented in the figure by a PARSEC
\citep{marigo+2017} isochrone with a very young age (40 Myr, red dotted
line). The top and right-hand sides are designed to exclude the bulk
of HB and RGB stars. In general, the location of the bottom side of
the box is somehow arbitrary, as there is no clear division between
the MS-TO and BSSs. Usually, it is assumed that the BSS population
begins 4-5 $\sigma$ magnitudes above the MS-TO,  where
  $\sigma$ is the photometric error at this magnitude level. However,  both to minimize the risk of
  contamination from MS-TO stars and blends, and to increase the
efficiency of the $A^+_{rh}$ parameter, \citet{ferraro+2018} selected only the brightest portion of the BSS sequence, where the most
massive BSSs are expected to be located. Thus, they considered only
BSSs that are approximately
$0.2 M_\odot$ more massive than MS-TO stars.
To construct the stellar evolutionary tracks in the F300X, we
downloaded PARSEC isochrones\footnote{http://stev.oapd.inaf.it/cgi-bin/cmd}  \citep{marigo+2017} in this filter in a very
wide age range (500 Myr-21 Gyr)  in steps of 10
Myr and with properties compatible with those of NGC 1835: metallicity
[Fe/H]$= -1.7$ and $\alpha-$element abundance [$\alpha$/Fe]$= +0.4$
\citet{mucciarelli+2021}, and standard helium abundance (Y = 0.248).
The track corresponding to the $0.8 M_\odot$ stars (rightmost solid
line in Fig. \ref{boxes}) well reproduces, as expected, the MS-TO
region. Hence, we adopted as threshold for the bright BSS sample the
position of the $1 M_\odot$ evolutionary track (leftmost solid line
in Fig. \ref{boxes}). The final BSS sample is shown by the blue
circles in the figure, and it contains 86 objects.

Since the MS-TO region is contaminated from LMC field stars, we chose
the RGB stars as reference population.  We selected them in the
optical
CMD (where they appear more evident and brighter than hot populations)
and in a magnitude range similar to that covered by the BSS sample in
this plane (see the black box in the right panel of
Fig. \ref{boxes}). This avoids both faint magnitudes, to prevent
incompleteness problems, and bright magnitudes, to prevent saturation
effects. We obtained a very rich final sample of 596 objects.
Given the large statistics of this sample, small differences in the
selection lead to negligible differences in the value of $A^+_{rh}$
(see below).

\subsection{The $A^+_{rh}$ parameter}
\label{sec:apiu}
As mentioned in Sec.\ref{intro}, the $A^+_{rh}$ parameter is defined
as the area enclosed between the normalized cumulative radial
distribution of BSSs and that of a reference population of lighter
stars, counted within one half-mass radius from the cluster centre.
To obtain a meaningful value of $A^+_{rh}$, it is therefore crucial to
assess the level of photometric completeness and of field
contamination of both population.

\begin{figure*}
     \centering
         \includegraphics[scale = 0.40]{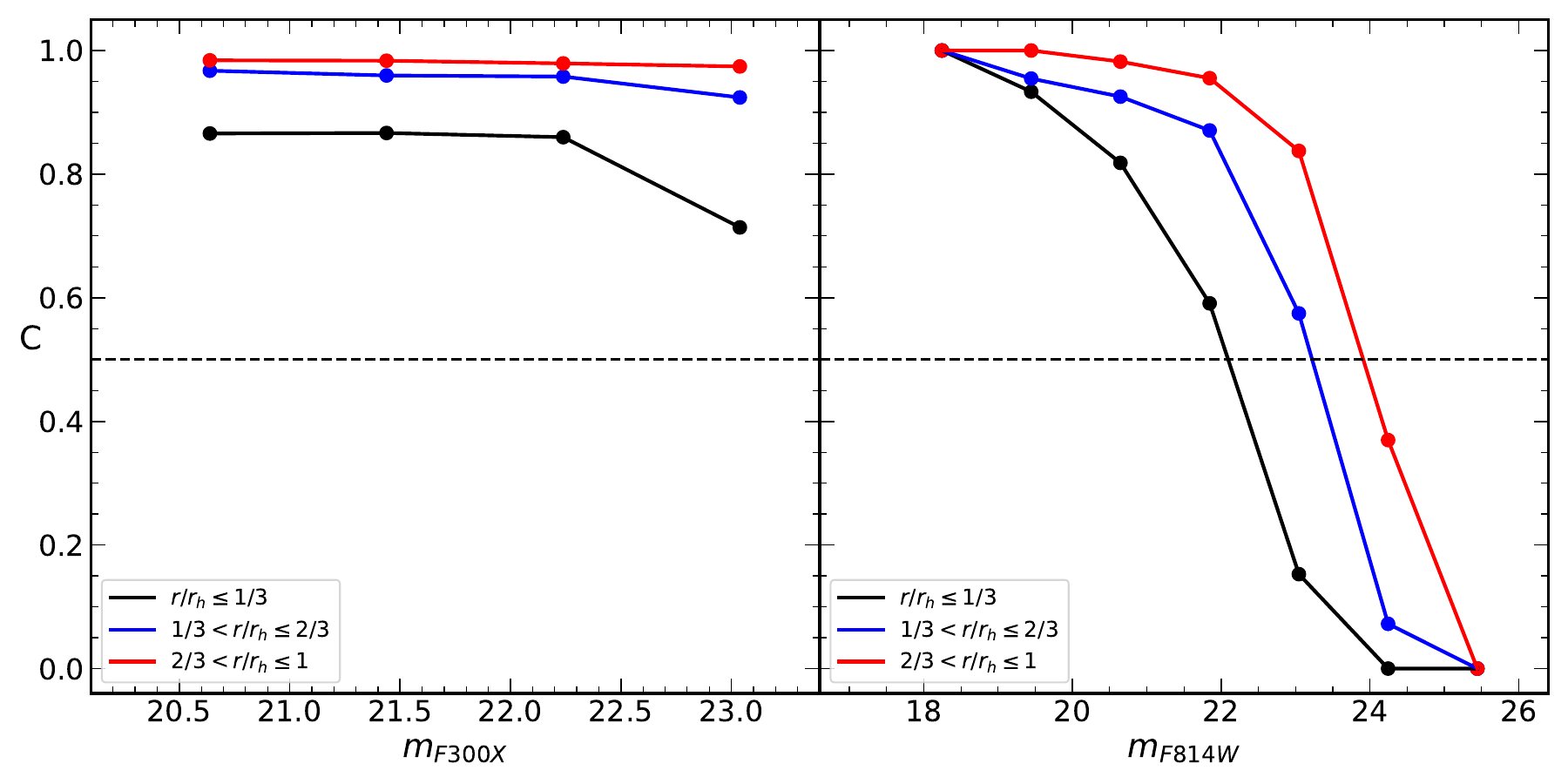}
         \caption{Completeness curves of the BSS and MS+RGB
           populations as a function of the $m_{\rm F300X}$ and the
           $m_{\rm F814W}$ magnitudes (left and right panels),
           respectively.  The different line colours refer to three
           different radial distances from the cluster centre (see
           labels).}
         \label{completezza}
\end{figure*}

To estimate the completeness information, we carried out artificial star
experiments using the prescriptions discussed in 
\citet{dalessandro+2015, cadelano+2020}.  We determined the mean ridge lines of the
BSS sequence (down to $m_{\rm F300X}\sim 23$) and of the RGB+MS
population (between $m_{\rm F814W} \sim 17.8$ and $m_{\rm F814W} \sim
26$) in the ($m_{\rm F606W}, m_{\rm F300X} -m_{\rm F606W})$ and
($m_{\rm F606W}, m_{\rm F606W}-m_{\rm F814W}$) CMDs, respectively.  We
then generated a list of artificial stars with an input $m_{\rm
  F606W}$ magnitude within the considered range, and assigned to each
star the corresponding ($m_{\rm F300X}-m_{\rm F606W}$) or ($m_{\rm
  F606W} -m_{\rm F814W}$) colour according to the respective mean
ridge line.  These artificial stars have then been added to the
acquired images using the DAOPHOT/ADDSTAR software. They have been positioned in a
grid of cells with side corresponding to ten times the typical FWHM of
the point spread function (PSF) and, in order to avoid artificial
crowding effects, only one star at a time was simulated in each
cell. Using the PSF model already derived, we repeated the data reduction process in the multiple modified images, and we
obtained a catalogue of $\sim 454000$ and $\sim 208100$ stars for the
BSS and the MS+RGB populations, respectively.  We simulated the stars
only for the WFC3 UVIS1 detector. This choice is dictated by the fact that the
$A^+_{rh}$ parameter is determined from the stars located within one
half-mass radius, which is just $13\arcsec$ for NGC 1835 (see
Sec. \ref{sec:densprof}).  The completeness C is defined as the ratio
between the number of stars recovered in output at the end of the
artificial star test (N$_o$), and the number of input stars actually
simulated (N$_i$). The completeness curves for the three radial bins
($r/r_h\leq 1/3$, $1/3 < r/r_h \leq 2/3$, $2/3 < r/r_h \leq 1$) that will be
used in the calculation of $A^+_{rh}$ (see Sec. \ref{BSS}) are shown
in Figure \ref{completezza}. The left panel shows the completeness
curves as a function of the $m_{\rm F300X}$ magnitude in the three
radial bins for the simulated BSSs, while the right panel shows C
versus $m_{\rm F814W}$ for the simulated MS+RGB population.  These
results guarantee that the selected BSS and RGB populations, which
extend down to $m_{\rm F300X}\sim 22$ and $m_{\rm F814W}\sim 20.2$
respectively, are poorly affected by incompleteness, both showing
C$>80\%$. Nevertheless, we corrected the two samples by using the
derived completeness curves: to build the cumulative radial
distributions, each star has been counted as 1/C (instead of 1), with
the value of C being derived from the incompleteness curves based on
the star magnitude and cluster-centric distance.  Thus the  
completeness-corrected total numbers of BSSs and RGB stars amount to
93.9 and 623.6 stars, respectively, with a correction of the observed sample of only a 
few percent ($9\%$ and $5\%$ respectively).  

\begin{figure}
     \centering
         \includegraphics[scale = 0.44]{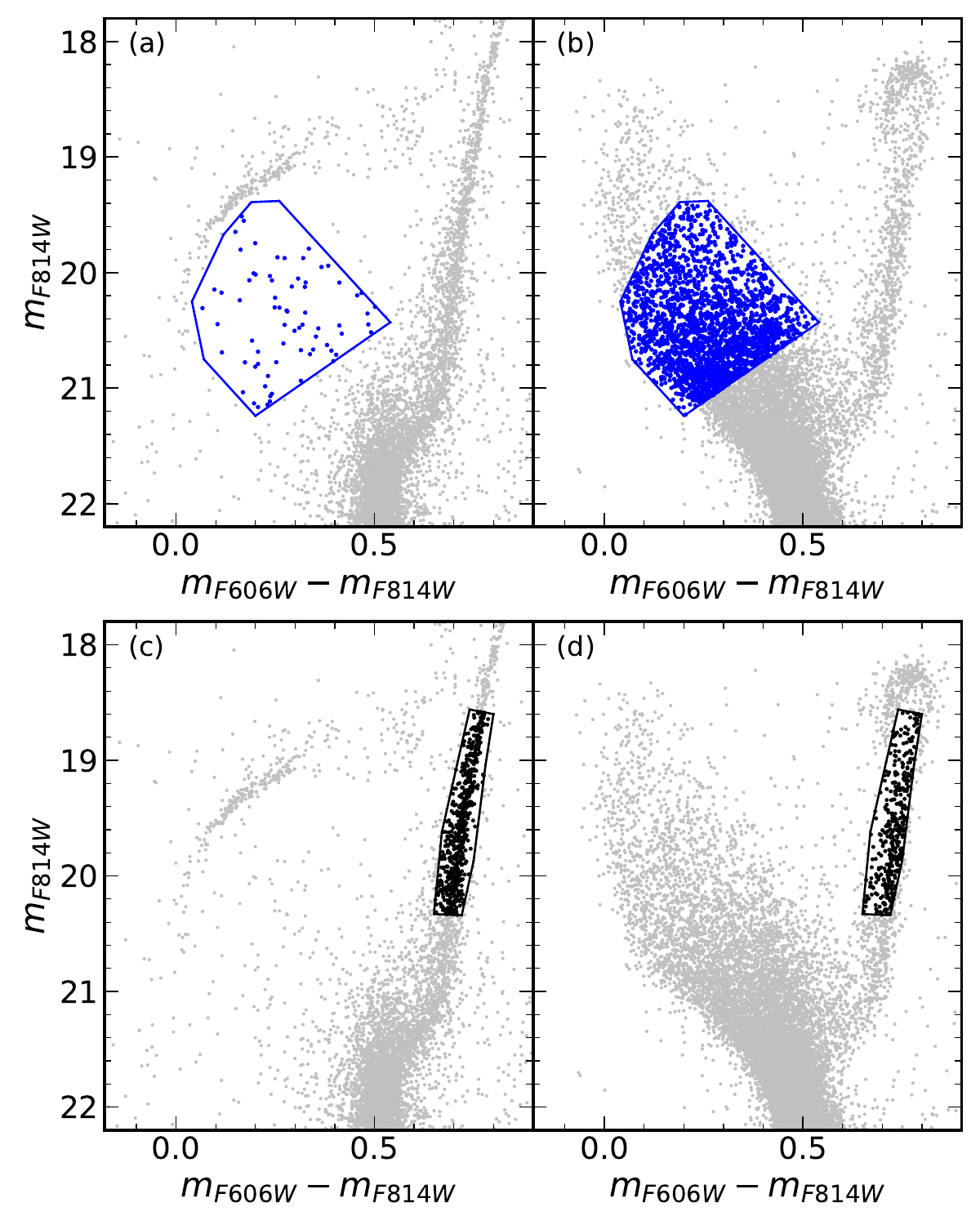}
         \caption{Decontamination process for the BSS and RGB samples. \emph{Panel (a)}: Optical CMD for the stars sampled
           by the WFC3 observations within one half-mass radius from
           the cluster centre (grey dots). The blue circles mark the
           position of the BSSs selected by using the near UV box
           drawn in Fig. \ref{boxes}, and the blue countour enclosing
           these objects is the adopted BSS selection box in the
           optical CMD.  \emph{Panel (b)}: Optical CMD of the stars
           sampled by the ACS parallel observations (grey dots). The
           blue contour is the optical selection box defined in panel
           (a), and the blue circles marks all the stars included
           within it.  \emph{Panel (c)}: Optical CMD for the WFC3
           stars sampled within one half-mass radius from the cluster
           centre (grey dots). The RGB selection is marked in black.
           \emph{Panel (d)}: Optical CMD of the ACS stars sampled
           (grey dots), with the same selection box marked in panel
           (c).  The black circles highlight the stars included within
           it.}
         \label{deco}
\end{figure}

To correct for LMC field contamination, we took advantage of the
parallel ACS observations.  Indeed, the CMD plotted in the rightmost
panel of Fig. \ref{cmds} shows that both the BSS and the RGB regions
suffer from a non negligible contamination from field stars, as
expected in the case of a GC located near the central bar of the LMC.
The decontamination via proper motions is not possible. In fact, just
another HST dataset centered on NGC 1835 is available in the archive,
but it has been acquired with the WFPC2 in 1995 \citep{olsen+1998},
while higher quality observations and a longer time baseline would be
necessary, given the distance of the LMC and the similar motion of the
GC and the host galaxy.  For this reason, we decided to perform a
statistical decontamination exploiting the parallel ACS
pointing \citep[e.g.][]{dalessandro+2019}. First of all, since for this dataset only the F606W and
F814W filters are available, we transported the BSS selection box
(previously defined in the near UV CMD; see Fig. \ref{boxes}) in the
optical ($m_{\rm F814W},m_{\rm F606W}$-$m_{\rm F814W}$) plane. This is
shown in panel (a) of Figure \ref{deco}, where the blue circles are
the BSSs selected in the near UV, and the box has been drawn to
enclose them. Panel (b) shows the same box superposed to the ACS
sample.  The selection box of RGB stars was already defined in the
optical plane, and it is shown superposed to the cluster WFC3 and the
field ACS CMDs in panels (c) and (d) of Fig.  \ref{deco},
respectively.  We counted the number of field stars falling in the two
boxes and we divided the values by the area sampled by the ACS/WFC
observations, obtaining the number density of field stars populating
the CMD regions covered by the two boxes: $\rho_{\rm field,BSS} =
0.0562 \pm 0.0012$ arcsec$^{-2}$, $\rho_{\rm field,RGB} = 0.0092 \pm
0.0005$ arcsec$^{-2}$.  Thus, we expect that within one half-mass
radius ($r_h=13\arcsec$), a total of 30 LMC stars contaminate the BSS
sample, and 5 contaminate the RGB sample.  However, the different
radial distribution of field stars with respect to that of cluster
stars must be taken into account for a proper field decontamination
procedure. In fact, while the distribution of the LMC field is
essentially uniform over the cluster size scale, the radial
distributions of cluster BSSs and RGB stars follow the King profile
(see Fig. \ref{profili}), with a sensibly increasing star density
toward the centre of the system.  Thus, we divided the field of view
included within $r_{h}$ in the same three concentric radial annuli
adopted above: $r/r_h \leq 1/3$, $1/3 < r/r_h \leq 2/3$, and $2/3< r/r_h \leq
1$.  Then, by multiplying the estimated star density by the
area of each radial bin ($A_{\rm bin}$), we determined the number of
contaminating field stars expected in the bin: $N_{\rm field,pop}(r) =
\rho_{\rm field,pop} \times A_{\rm bin}$ (with pop=BSS, RGB).  The
resulting values are listed in Table \ref{table:2}.

\begin{table*}
\renewcommand*{\arraystretch}{2.0}
\centering 
\begin{tabular}{c c c c c}
\hline
& N$_{\rm TOT}$ & N($r/r_h<1/3$) & N($1/3< r/r_h < 2/3$) & N($2/3 < r/r_h < 1$) \\
\hline
BSS & 93.9 (30) & 48.6(3) & 25(10) & 20.3(17)\\
RGB & 623.6 (5) & 165.6(1) & 280.2(1) & 177.8(3) \\
\hline
\end{tabular}
\caption{Completeness-corrected samples of BSSs and RGB stars (top and
  bottom row, respectively), and number of estimated field
  contaminants (values enclosed in the brackets). N$_{\rm TOT}$ is the
  total number of stars, while the last three columns provide the
  numbers of stars estimated in the three adopted radial bins.  }
\label{table:2}
\end{table*}

\begin{figure}
     \centering
         \includegraphics[scale = 0.25]{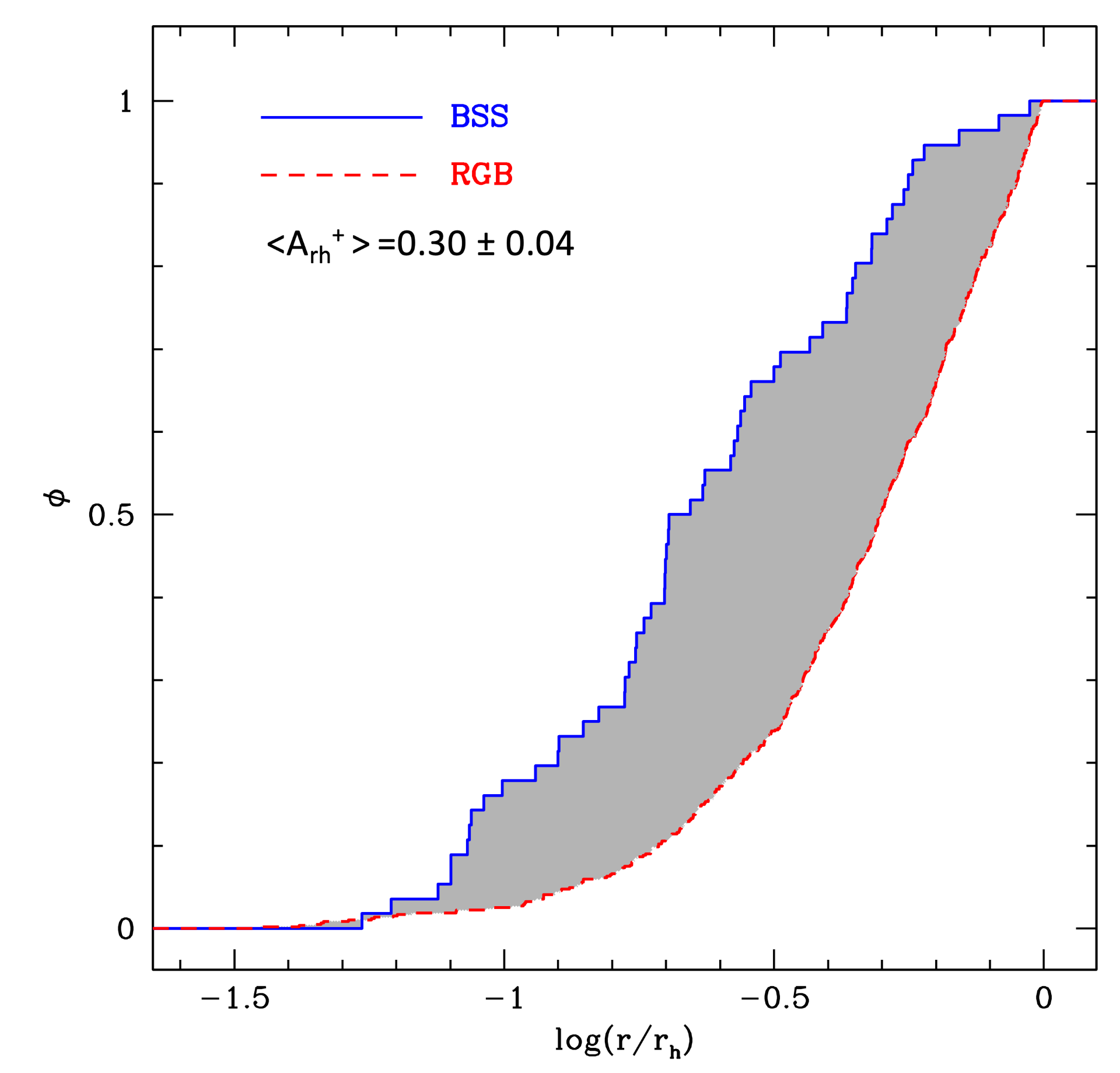}
         \caption{Normalized cumulative radial distributions of BSSs
           (blue solid line) and RGB stars (red dashed line) in one of
           the random realizations used to estimate the $A^+_{rh}$
           parameter in NGC 1835. The area enclosed between the two
           curves corresponds to the value of $A^+_{rh}$ (see
           label).}
         \label{apiu}
\end{figure}

To build the cumulative radial distributions, we randomly subtracted
these numbers of stars from the completeness-corrected samples of BSSs
and RGB stars, and we obtained the value of $A^+_{rh}$ as the area of
the region bounded by the two cumulative functions.  We repeated the
measurement of $A^+_{rh}$ 100 times, each time randomly removing from
the samples a number of stars equal to $N_{\rm field,pop}(r)$ in each
bin. We then adopted as final estimate of the parameter the average of
the 100 values of $A^+_{rh}$ computed in this way, finding
$<A^+_{rh}>=0.30$. 
Once the photometric incompleteness and field contamination are taken
into account, the main uncertainty affecting the result is due to the
limited statistics of the BSS sample. We therefore evaluated it by
using a jackknife bootstrapping technique (see \citealt{lupton+1993}).  Given
a sample of $N$ BSSs, it consists in repeating the calculation of
$A^+_{rh}$ for $N$ times, each time removing one different star.  The
final uncertainty is therefore calculated as $\sigma_{A^+_{rh}} =
\sigma_{\rm distr} \times \sqrt{(N-1)}$, where $\sigma_{\rm distr}$ is
the standard deviation of the $N$ estimates of the $A^+_{rh}$.  We
found $\sigma_{A+} = \pm 0.04$.  One of the random realizations of the
cumulative radial distributions of BSSs (blue curve) and RGB stars (in
red) is shown in Figure \ref{apiu}, where the grey-shaded area
represents the $A^+_{rh}$ parameter.

\section{Discussion and Conclusions}
\label{disc}
\begin{figure*}
     \centering
         \includegraphics[scale = 0.45]{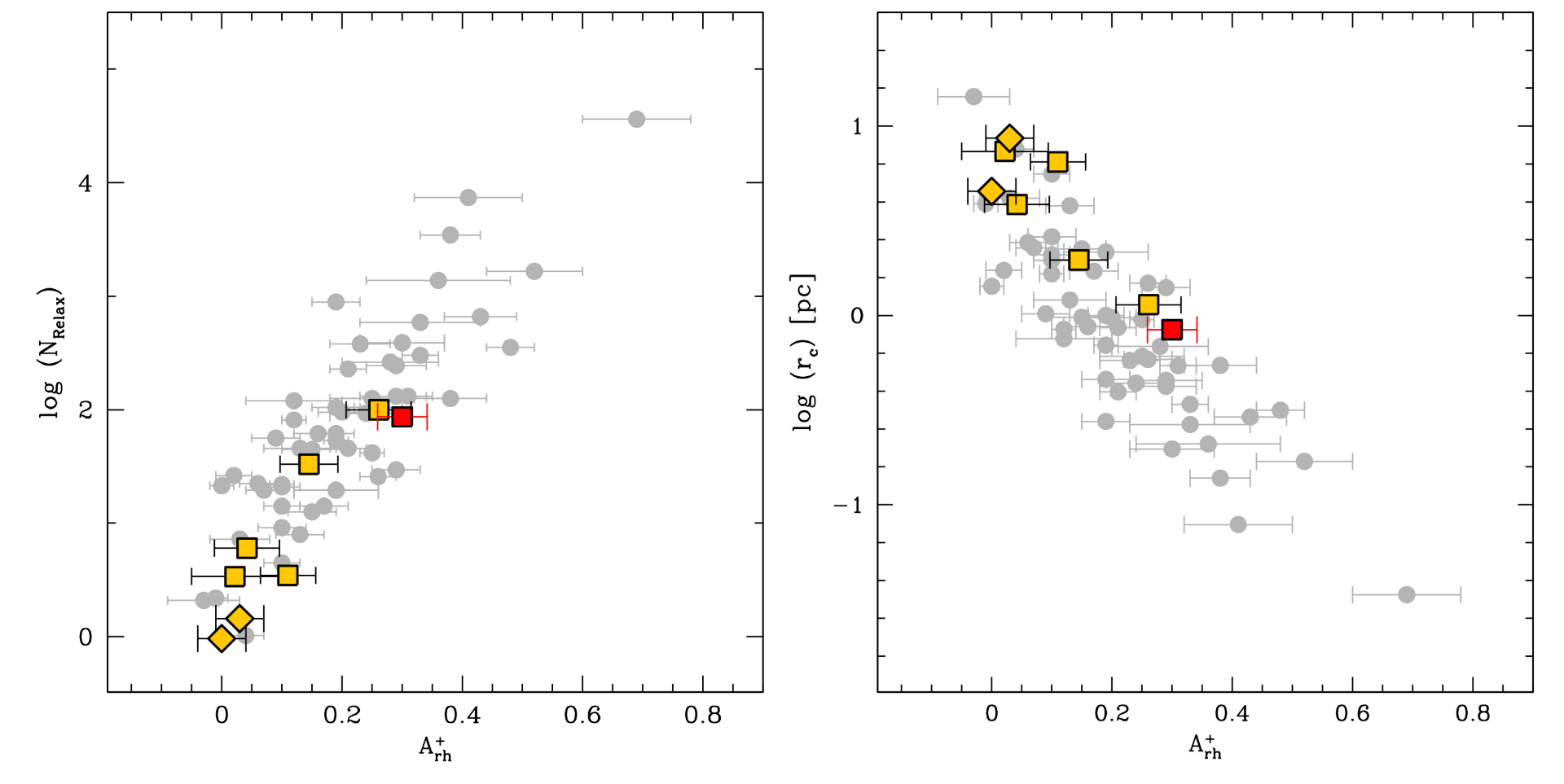}
         \caption{Final results for NGC 1835 in comparison with other clusters. \emph{Left panel}: Relation between $N_{\rm relax}$
           and $A^+_{rh}$ for the star clusters homogeneously
           investigated so far through the dynamical clock approach:
           the 52 Galactic GCs discussed in \citet{ferraro+2023} are
           shown as grey circle, the five LMC clusters discussed in
           \citet{ferraro+2019} are plotted as yellow squares, the two
           young star clusters in the SMC discussed in
           \citet{dresbach+2022} are marked as yellow diamonds, and the
           determination of NGC 1835 obtained in this work is plotted
           as a filled red square.  \emph{Right panel}: Relation
           between $r_c$ and $A^+_{rh}$ for the same star clusters
           plotted in the left panel (same symbols). }
         \label{conundrum}
\end{figure*}

The result obtained for NGC 1835 can now be compared with the outcomes
of previous works. Indeed, similar studies were conducted for a total
of 52 Galactic GCs (48 systems in \citealt{ferraro+2018}, NGC 6256 in
\citealt{cadelano+2022b}, and 3 other clusters in
\citealt{ferraro+2023}), plus a total of 7 clusters in the Magellanic
Clouds (5 old systems in the LMC and two young ones in the SMC; see
\citealt{ferraro+2019} and \citealt{dresbach+2022}, respectively).
The left panel of Figure \ref{conundrum} shows the trend between
$N_{\rm relax}$ (the number of central relaxation times suffered by
each system) and the value of $A^+_{rh}$ for all the clusters
consistently investigated in previous studies and for NGC 1835 (shown
in red).  The new determination well fits into, and therefore further
reinforce, the close correlation observed between the two parameters.

It is interesting to point out that the MW, the LMC and the SMC clusters all fall along the same relationship, indicating that the dynamical clock parameter can provide a measure of cluster’s dynamical ages also outside the MW. As previously discussed, the existence of such a strong correlation between $A^+_{rh}$ and $N_{relax}$ clearly shows that they both trace the dynamical aging of star clusters. However, \citealt{ferraro+2023} pointed out that $A^+_{rh}$ provides a better measure of the dynamical history and dynamical age of a cluster than $N_{relax}$.  As discussed in \citet{ferraro+2023}, $t_{rc}$ is a parameter determined by the present-day dynamical properties of star clusters, which can be the result of quite different combinations of initial conditions and evolutionary paths. Hence, $t_{rc}$ is not necessarily
representative of a specific and unique dynamical history of each stellar system. Conversely, the current level of BSS central concentration is the direct outcome of the many internal and external phenomena that, during the entire cluster lifetime, have modified and characterized the mass segregation process. Thus, the main advantages of the
$A^+_{rh}$ parameter with respect to $t_{rc}$ can be summarized as
follows: (1) the $A^+_{rh}$ parameter is entirely and directly
measured from observations; (2) it is specifically designed to have
maximum sensitivity in quantifying the mass segregation of the heaviest
and brightest (hence, easily identifiable) stellar population (the
BSSs); (3) at odds with $t_{rc}$, its definition does not require
over-simplying assumptions (e.g. spherical symmetry, no rotation,
King model structure) or approximations (for instance the definition
of $r_c$, which is very uncertain, in particular, in dynamically evolved stellar system
where the presence of a central cusp makes the definition more complicated and possibly meaningless).

In addition, the value of $A^+_{rh}$ seems to be a very promising
discriminator between non core-collapsed and core-collapsed
clusters. In fact, \citet{ferraro+2023} have shown that 7 Galactic GCs
with measured $A^+_{rh}$ and classified as post core-collapsed
clusters from the evidence of a central cusp in their density profile
all exhibit $A^+_{rh}> 0.29$. In this respect, NGC 1835 (for which we
estimate $A^+_{rh} = 0.30\pm 0.04$) likely is on the verge
to core collapse.

The right panel of Fig. \ref{conundrum} shows the relationship between
$A^+_{rh}$ and the core radius for the same systems. As already
discussed in \citet{ferraro+2019}, it shows that clusters with a larger core radius correspond to dynamically younger systems characterized by low degrees of spatial segregation of their most massive visible stars (corresponding to lower $A^+_{rh}$ values) while those with a smaller core radius are in more advanced dynamical stages with a stronger degree of mass segregation. NGC 1835 nicely falls along this anticorrelation, showing the highest value of $A^+_{rh}$ in the LMC (hence, the most advanced dynamical age), consistent with its very small core radius value. 
So, while \citet{mackey+2008} proposed to resolve the size-age
conundrum by advocating the action of binary BHs that progressively
drive core expansion of the systems, the range of clusters core radii may simply be a consequence of differences in the clusters dynamical ages. Furthermore, as discussed by \citet{ferraro+2019}, young and old clusters exhibit very different properties in terms of mass (where the older clusters are more massive than the younger ones) and distance from the LMC centre, thus excluding the existence of a simple and direct evolutionary sequence linking the two groups of clusters.
 
Finally we point out that the young clusters in the LMC are low-mass systems with small core radii and located in the most central regions of the galaxy (see the middle and bottom panels in Fig.1 of \citet{ferraro+2019}).
The lack of young clusters with large core radii may be due to the early disruption of these clusters; as proposed by \citet{ferraro+2019}, it is plausible that, in this low-mass regime, only the most compact clusters managed to endure the tidal forces exerted by the host galaxy and survived to the present days, while the low mass/low-concentration systems have been disrupted.

In conclusion, the study of the structure of NGC 1835 and the degree of segregation of its BSS presented in this paper further extended previous investigations by adding information about a dynamically old cluster characterized by a significant BSS spatial segregation and bridging the gap between the dynamical characterization of clusters in the Milky Way and the Magellanic Clouds.

\begin{acknowledgements}
This work is part of the project Cosmic-Lab at the Physics and
Astronomy Department ``A. Righi'' of the Bologna University (http://
www.cosmic-lab.eu/ Cosmic-Lab/Home.html). 
\end{acknowledgements}

%
%

\end{document}